\documentclass[12pt]{article}

\newcommand\re[1]{{(\ref{#1})}}

\newcommand{\be}{\begin{equation}}
\newcommand{\ee}{\end{equation}}
\newcommand{\bea}{\begin{equationarray}}
\newcommand{\eea}{\end{equationarray}}
\newcommand{\ba}{\begin{array}{c}}
\newcommand{\ea}{\end{array}}

\newcommand{\de}{\delta}
\newcommand{\e}{\epsilon}
\newcommand{\ve}{\varepsilon}
\newcommand{\vp}{\varphi}
\newcommand{\pa}{\partial}
\newcommand{\fr}{\frac{1}{2}}
\newcommand{\na}{\nabla}

\newcommand{\La}{\Lambda}

\begin{document}

\thispagestyle{empty}

\begin{flushright}
\vspace{1mm} hep-th/0103028                                                                                                                            \\
FIAN/TD/32/01\\
March 2001\\
\end{flushright}

\vspace{10mm}

\begin{center}
{\bf A generating formulation  \\for free higher spin massless fields } \\

\vspace{10mm}
Arkady Yu. Segal ${}^{(1)}$
\\

\vspace{1cm}   {\it
I.E.Tamm Department of Theoretical Physics, Lebedev Physics
Institute,\\
Leninsky prospect 53, 117924, Moscow, Russia\\
e-mail: segal@lpi.ru  }

\vspace{2mm}

\end{center}
\vspace{5mm}
\begin{abstract}
An action describing the dynamics of
an infinite collection of massless integer spin fields with spin
$s=0,1,2,3, ...,\infty$ corresponding to totally symmetric Young
tableaux representations of Poincare and anti-de Sitter groups is
constructed, in any dimension $d$, in terms of two functions on a
$2d$-dimensional manifold. The action is represented by an integral
localized on a $2d-1$-dimensional hypersurface.
\end{abstract} \vspace{1cm}

\vspace{6cm}
--------------------------------------

{\tt $(1)$ on leave of absence from}

Department of Physics, Tomsk State University, Russia.

\newpage
In this paper, we construct a generating model describing the Euclidean
version of a sum of the actions for integer spin massless fields
(Fronsdal fields),
with spin range from $0$ to $\infty$, every spin enters one
time.

The free integer spin-$s$ massless field models corresponding to totally
symmetric (one row Young tableaux) representations are formulated, in a
space-time of any dimension $d$ and any signature,
equipped with the metric $g_{mn}(x)$
of constant curvature
$R_{mnpq}(x)=\Lambda (g_{mp}\, g_{nq} -g_{np}\, g_{mq}),\, \Lambda =const$, in terms of a rank-$s$ tensor field
$\vp_{m(s)}(x)$ subject to the double-tracelessness constraint \be
\label{dtrace} {{\vp_{kl}}^{kl}}_{m_5...m_s} =0 , \ee where the
contraction is performed by the background metric.  For $s=0,1,2,3$
this constraint is satisfied automatically.
Equivalently, $\vp_{m(s)}$\footnote{
We use the following condensed notation for symmetric tensors:
the expression of the type
$$
X^{n(k)} Y^{n(l)} Z^{m(n)}
$$
denotes the totally symmetrized tensor of rank $k+l+m$ obtained by
a symmetrized product of three totally symmetric tensors of ranks $k,l,m$.
The symmetrization operation is a projector, so that
$
X^{n} Y^{n} \mapsto \frac{1}{2}(X^{n_1} Y^{n_2} +X^{n_2} Y^{n_1})
$
etc. All indices are raised and lowered by the background metric.}
may be represented as a sum of its
traceless $\phi_{m(s)}$ and trace $\chi_{m(s-2)}$ parts
\be \label{1}
\vp_{m(s)} =\phi_{m(s)} + g_{m(2)} \chi_{m(s-2)}\,,\,
{{\phi_k}^{k}}_{m(s-2)} = 0\,,\, {{\chi_k}^{k}}_{m(s-4)} = 0.
\ee
The gauge
transformations are
\be \label{frlaw} \delta \vp_{m_1...m_s}
=\nabla_{(m_1} \ve_{m_2...m_s)} \ee where $\nabla_m$ is the covariant
derivative constructed out of the metric and $\ve_{m(s-1)}$ is
traceless, \be \label{etrace} {\ve^{k}}_{km_3...m_{s-1}} =0.  \ee For
$s=1,2$, this constraint is automatically fulfilled.
Equivalently, the gauge transformations are rewritten as
\be \label{frlaw1} \ba
\delta \phi_{m(s)}
={\mbox{Traceless part of}}\; \nabla_{m} \ve_{m(s-1)}\\ \\
\delta \chi_{m(s-2)}
= \frac{s-1}{s-2+d} \nabla^n {\ve_{nm(s-2)}}  \ea \ee
These transformations are known to determine the action ${\cal
A}_s[\vp_s]$ unambiguously, provided the condition of the absence of
higher derivatives is imposed. In the flat case ($\Lambda=0$), the
action has the form \cite{fronsdal1},
\be  \label{fraction}
\begin{array}{cllc}
{\cal A}_s[\vp_s]=\fr (-)^s \int d^d x \bigl\{
\pa_{n} \vp_{m_1...m_s} \;\pa^{n}  \vp^{m_1...m_s}& &\\ & &\\
-\fr s(s-1) \pa_{n} {\vp^{k}}_{km_1...m_{s-2}}\; \pa^{n}{{\vp^{k}}_k}^
{m_1...m_{s-2}} & &\\ & &\\
+ s(s-1) \pa_{n} {\vp^{k}}_{km_1...m_{s-2}}\; \pa_{l} \vp^{nlm_1...m_{s-2}}
-s \,\pa_{n} {\vp^{n}}_{m_1...m_{s-1}}\; \pa_{k} \vp^{km_1...m_{s-1}}& & \\ & &\\
-\frac{1}{4} s(s-1)(s-2)
\pa_{n} {{\vp^{k}}_k}_{nm_1...m_{s-3}}\; \pa_{r}
{{\vp^{l}}_l}^{rm_1...m_{s-3}}\bigr\} ,& &
\end{array}
\ee
where all contractions are performed by the flat metric
$g_{mn}=\eta_{mn}$. For $s=0$ there are no gauge parameters at all,
i.e.  a theory is not a gauge one.  Nevertheless, the formulas
\re{fraction} make sense in this case and provide the massless scalar
theory.
Though the actions \re{fraction} were originally constructed  in $4D$
\cite{fronsdal1} they do not contain explicitly the dimension $d$ of
spacetime and describe consistently massless higher spin dynamics for any $d$.
We shall refer to them as Fronsdal actions.

For the nonzero constant curvature case $\La \neq 0$ which we will refer to
as $AdS$ case\footnote{In this paper we do not consider supersymmetry
and unitarity questions, therefore the sign of $\La$ is inessential.}, the gauge invariant action is
obtained after substituting $\eta_{mn} \mapsto g_{mn}\,,\,\pa_m
\mapsto \nabla_m$ and adding some mass-like terms, proportional to
$\Lambda$.  As a result, the gauge-invariant actions describing
massless fields of arbitrary spin exist in the constant curvature
backgrounds \cite{fronsdal1}. However, if one takes a general metric background, the
{\it higher spin problem} arises as the gauge variation of the action
\re{fraction} (or its $AdS$ deformation) appears to be proportional
to the total Riemann curvature
$R_{mnpq}(x)-\Lambda (g_{mp} g_{nq} -g_{np} g_{mq})$ which is
impossible to cancel by any modifications of the gauge laws
\re{frlaw} and by adding nonminimal terms.
This inconsistency is a core of the higher spin problem
\cite{ardeser}, \cite{Weinberg:1980kq}.
The same problem of the gauge invariance breakdown
virtually appears when one tries to introduce interactions between
the higher spin fields.
The significant progress in overcoming the problem had been achieved
by Fradkin and Vasiliev \cite{vas1} and later on by Vasiliev
\cite{vas2}, see \cite{vasO} for the review. A general message from
their works is that

1) any consistent theory of interacting higher spin gauge
fields (including gravity) should contain an infinite number of
fields with infinitely increasing spin

2) a natural vacuum in the gravitational sector  of higher
spin models is $AdS$ space, not the Minkowski one. Interaction is nonanalytic in the cosmological constant
and contains higher derivatives up to an infinite order.

The major success of the program undertaken in a series of papers
\cite{vas1},\cite{vas2},\cite{vas234}, based on the "unfolded formulation"
technology,
is the construction of the consistent equations of motion
for $4d$ interacting higher spin fields.
By now, despite the consistent $d=2,3,4$ equations of motion are
constructed \cite{vasO} and an action principle is suggested for $d=3$ theory
\cite{vas3s}, the higher spin actions are not known for $d>3$ and
even the higher spin equations are not available for $d>4$.
This circumstance, though exhibiting neither dead end nor stagnation
the unfolded formulation approach to the higher spin problem, prompts one
to develop alternative approaches
to the higher spin problem.

Here we construct a simple
generating formulation of the higher spin dynamics in $AdS_d$
space (boson case). We hope this formulation may serve as a good
starting point for studying the interaction problem. In this paper,
we will present the Euclidean version of the theory, so the $AdS_d$
space is understood as an open ball in a $d$-dimensional Euclidean
space with the metric of a constant negative curvature (see, e.g.
\cite{Witten:1998qj}).
Consider the cotangent bundle to $AdS_d$, $T^* (AdS_d)$, parametrized
by the coordinates  $x^m$ and momenta $p_n$. The higher spin fields
are described by two "functions" $h_1(x,p)$ and $h_2(x,p)$  which
have a form of formal power series in momenta:
\be \label{hclass}
h_{1,2} = \sum\limits_{k=0}^\infty h_{1,2}^{m(k)} (x)
p_{m_1}...p_{m_k}. \ee
Any function $F(x,p)$ in this paper is also considered as a power
series in momenta.

\noindent Introduce the "Hamiltonian" function
\be H(x,p)=g^{mn}(x) p_m p_n \ee and
the "covariant derivative"
\be \nabla_m =\pa_m + \Gamma^k_{mn} (x)
p_k \de^n\,,\, \pa_m\equiv \frac{\pa}{\pa x^m}\,,\, \de^n \equiv
\frac{\pa}{\pa p_n},
\ee
where $\Gamma^k_{mn} (x)$ is a Riemannian
connection corresponding to the metric $g_{mn}(x)$, so that
\be \ba
\nabla_m
\sum\limits_{k=0}^\infty F^{m_1...m_k} (x) p_{m_1}...p_{m_k} =
\sum\limits_{k=0}^\infty (\nabla_m
F^{m_1...m_k} (x)) p_{m_1}...p_{m_k},\\ \\ \nabla_m (A(x,p)B(x,p)) =
(\nabla_m A)B +A(\nabla_m B),\\ \\ \nabla_m H(x,p)=0 \\ \\
{[\nabla_m, \nabla_m]} F(x,p) = p_k\,{{R_{mn}}^k}_l (x)\, \de^l  F(x,p)=
\La \,(p_m \de_n -p_n \de_m) \, F(x,p).  \ea \ee

\noindent The gauge transformations are
\be \label{Glaws}
\ba
\delta h_1 =  \, (H-\mu^2) \, a + p^m \nabla_m \,\e ,\\ \\
\delta h_2 =  \, (H-\mu^2) \, c+ 2\,a  + \delta^m \nabla_m \,\e,
\ea
\ee
here $\e,a,c,$ are arbitrary smooth functions having a compact support
in the $x$-space. $\mu^2 \neq 0$ is an arbitrary dimensionless constant,
the models with different $\mu^2$ appear to be identified by
purely algebraic field redefinition.
Let us show that the transformations \re{Glaws} provide the correct
gauge laws for the infinite  collection of $AdS_d$ massless fields
with integer spins from 0 to $\infty$. To this end we need some
simple tools to handle traces of tensor coefficients of arbitrary functions.
Let us note that,
given any function $$F(x,p)=\sum\limits_{k=0}^\infty F^{m(k)} (x)
p_{m_1}...p_{m_k},$$ one can unambiguously represent it in the
form
\be \label{decom} F(x,p)=
\sum\limits_{l=0}^\infty \sum\limits_{k=0}^\infty
F_{(l)}^{m(k)} H^l p_{m_1}...p_{m_k},
\ee where $F_{(l)}^{m(k)}$ are
traceless, ${{F_{(l)}}{}_n}^{nm(k-2)}=0$. This is easily done by
decomposing each $F^{m(k)}$ to its traceless part and the traces
$F^{m(k)}=F_{(0)}^{m(k)}+g^{m(2)}F_{(1)}^{m(k-2)} +g^{m(2)}\, g^{m(2)}
F_{(2)}^{m(k-4)} +...$, then summing up the power series by momenta and
noting that the trace parts give the powers of $H$.
The decomposition \re{decom} is then rewritten as
\be
F(x,p)=
\sum\limits_{k=0}^\infty
F^{m(k)} (H) p_{m_1}...p_{m_k},
\ee
where $F^{m(k)} (H)=
\sum\limits_{l=0}^\infty
F_{(l)}^{m(k)} H^l. $
Decomposing the power series $F^{m(k)} (\sigma)$ at the point
$\sigma=\mu^2$ one gets
\be \label{decom2}
\ba
F(x,p)=
\sum\limits_{k=0}^\infty
F_{[\mu^2]}^{m(k)} (H-\mu^2) p_{m_1}...p_{m_k}=
\sum\limits_{l=0}^\infty \sum\limits_{k=0}^\infty
F_{[\mu^2](l)}^{m(k)} (H-\mu^2)^l p_{m_1}...p_{m_k}=\\ \\
\sum\limits_{l=0}^\infty
F_{[\mu^2](l)}(H-\mu^2)^l,
\ea
\ee
where the power series $F_{[\mu^2](l)}$ contain only traceless
coefficients. Given $\mu^2$, we will say that the
$F_{[\mu^2](0)}$ term is the traceless part of the function $F(x,p)$
and the first, second and further traces of $F$ are represented by
$F_{[\mu^2](l)} (H-\mu^2)^l$ with $l=1,2,...$ forming altogether the
traceful part of $F$. The function is traceless if
it is equal to its traceless part. In this sense each coefficient
$F_{[\mu^2](l)}$ is a traceless function.

Now represent all the entries of the gauge transformation laws
\re{Glaws} in the form \re{decom2} to get
\be
\label{Glaws2}
\de \sum\limits_{l=0}^{\infty}  h_{1[\mu^2]\,(l)}\, (H-\mu^2)^l =
\sum\limits_{l=0}^{\infty} \left\{a_{[\mu^2]\,(l)} (H-\mu^2)^{l+1} +
p^m \nabla_m
\e_{[\mu^2]\,(l)} (H-\mu^2)^l\right\}
\ee
$$
\de \sum\limits_{l=0}^{\infty}  h_{2[\mu^2]\,(l)} (H-\mu^2)^l =
\sum\limits_{l=0}^{\infty} \left\{c_{[\mu^2]\,(l)} (H-\mu^2)^{l+1} +
2  a_{[\mu^2]\,(l)} (H-\mu^2)^l +
\de^m \nabla_m
\e_{[\mu^2]\,(l)} (H-\mu^2)^l \right\},
$$
wherefrom it is seen that all the traces of $h_1$ and $h_2$ may be
gauged away by $a$ and $c$-transformations. In fact, the
destination of $a$ and $c$ transformations is to gauge away the
traces of $h_{1,2}$.
It is worth noting that the
traceful part of $\e$ is already contained in $a$ and $c$ as the
gauge transformations \re{Glaws2} do not change if one redefines
$\e,a,c$ according to
\be
\de \e= (H-\mu^2)\nu\,,\,\de a= - p^m \na_m \nu\,,\,
\de c=\de^m \na_m \nu.
\ee
Therefore without loosing a generality one may set $\e$
traceless
\be
\e= \e_{[\mu^2]0} \equiv \ve =\sum\limits_{k=0}^{\infty}
\ve^{m(k)} p_{m_1}...p_{m_k}\,;\, {\ve_n}^{nm(k-2)}=0.
\ee
Now suppose an action
${\cal A}[h_1, h_2]$ invariant w.r.t.  gauge transformations \re{Glaws},\re{Glaws2}
exists, then as far as $a$ and $c$-transformations are purely
algebraic, $h_1$ and $h_2$ should enter ${\cal A}[h_1, h_2]$ in
$a$ and $c$-invariant combinations only. It is easy to see that
there are only two $a,c$-invariants of the form
\be  \label{5}
f_1 =h_{1[\mu^2](0)}\,,\,f_2 =h_{1[\mu^2](1)}-\frac{1}{2}
h_{2[\mu^2](0)}, \ee parametrized by two traceless functions $f_1,
f_2$.  It is easy to derive the $\ve$ transformation laws for the
coefficients of $f_1, f_2$ which read
\be \label{3}
\ba
\de f_{1\,m(s)} =( {\mbox{Traceless part of}}\; \nabla_{m} \ve_{m(s-1)}) +\mu^2
\,
\frac{s+1}{s+d}\, \nabla^n {\ve_{nm(s)}}\\ \\
\de f_{2\, m(s)} = \frac{(2-2s-d)(s+1)}{2(2s+d)}\, \nabla^n {\ve_{nm(s)}}.
\ea
\ee
Making the linear change  of variables
\be \label{2}
\phi_{m(s)}=f_{1\,m(s)} -\mu^2 \frac{4s+2d }{2-2s-d} \,f_{2\,m(s)}\;,\;
\chi_{m(s)}= \frac{4s+2d }{2-2s-d} \,f_{2\,m(s)},
\ee
one gets
\be \label{frlaw2} \ba
\delta \phi_{m(s)}
={\mbox{Traceless part of}}\;\; \nabla_{m} \ve_{m(s-1)}\\ \\
\delta \chi_{m(s-2)}
= \frac{s-1}{s-2+d} \nabla^n {\ve_{nm(s-2)}}  \ea \ee
which is nothing but the gauge laws for Fronsdal fields \re{frlaw1}.
Therefore, the fields $f_1, f_2$ present the infinity of
Fronsdal double traceless higher-spin fields and the gauge
transformations \re{Glaws} present a generating formulation of
gauge laws equivalent to Fronsdal ones.

Before turning to the construction of an invariant action in
generating form it is worth discussing some algebraic properties of
the operators governing the gauge laws \re{Glaws}. There are 3
operators
\be
A=\de^m \na_m \,,\, B=p^m \na_m\,,\, H=p^2,
\ee
satisfying the commutation relations
\be \label{kinem}
[A,H]=2B \,,\, [H,B]=0
\ee
and
\be \label{ncym}
{[B, [B,A] ]} =V_{11} B +V_{12} A\,,\,{[A,[A,B] ]}=V_{21} B
+V_{22} A,
\ee
where
\be \ba
V=\left( \begin{array}{cc} V_{11} & V_{12} \\ V_{21} & V_{22} \ea\right) =2 \La \left( \begin{array}{cc} 2p_m \de^m +d+3 & -2 p^2 \\
-2 \de^2 & 2p_m \de^m +d-3 \ea \right). \ea \ee
$V$ does not contain $x$-derivatives, this set of operators
is the only place $\La$ explicitly enters all the expressions below.

Now let us unite $h_1$ and $h_2$ in a two-component column
\be
h=\left(\ba h_1\\h_2 \ea\right)
\ee
so that the gauge transformations
\re{Glaws} read
\be
\de h = \left(\ba B\\ A \ea\right)\e +\left(\ba H-\mu^2 \\ 2
\ea\right) a + \left(\ba 0\\ H-\mu^2 \ea\right) c,
\ee
and introduce the "almost wave
operator" $N$
\be \label{alm}
N h = \left( \begin{array}{cc} N_{11} & N_{12} \\ N_{21} & N_{22} \ea\right)=
\left\{ \left( \begin{array}{cc} AB - 2
BA & B^2 \\ A^2 & BA-2AB \ea\right) - V\right\} h.\ee
This operator projects out $\e$-transformations, as
\be
N \left(\ba B\\ A \ea\right) =
\left(\ba {[B,[B,A] ]}\\ {[A, [A,B] ]}\ea\right) -V\left(\ba B\\ A \ea\right)
\ee
that is zero by virtue of the relations \re{ncym}.
To construct the completely invariant wave equation one has to
find the $a,c$-invariant combination of the components of
$N h$.

Before presenting the equation it is worth to exhibit the structures
ensuring the construction of the action. We need an integration
procedure over $T^*(AdS_d)$ which should allow us to construct
meaningful expressions for an action and to reexpress them in terms
of components fields, i.e. in terms of tensor fields over $AdS_d$.
It appears that such an integration procedure does exist and the
measure of integration is provided by a distribution localized on the
surface $H-\mu^2=0$. Introduce the
"step" $\theta$-function $\theta(\sigma)$ :
\be
\ba
\theta(\sigma)=0\;\;,\;\; \sigma < 0 \\ \\
\theta(\sigma)=\fr\;\;,\;\; \sigma = 0\\ \\
\theta(\sigma)=1\;\;,\;\; \sigma > 0.
\ea
\ee
and consider the integral
\be \label{bint}
I_D (x) = \int d^d p \;\theta(-H+D),
\ee
where $D=\mu^2 >0$. As the metric $g_{mn}(x)$ is Euclidean, for
every point $x$ of $AdS_d$ the equation $H=D$ defines a
$d-1$-dimensional ellipsoid in the $p$-space, and $-H+\mu^2 >0$ inside
the ellipsoid while $-H+\mu^2 <0$ outside. Therefore the integration
in \re{bint} goes over the interior of the ellipsoid and the
integral is easily calculated to be
\be \label{bint2}
I_D = \mu^d\, b_{(d)}\,\,\sqrt{det(g_{mn}(x))} ,
\ee
where $b_{(d)}$ is the volume of a $d$-dimensional unit ball.
Analogously, the integrals of the form
\be \label{bint2}
I_D [F] = \int d^d p \;\theta(-H+D) F(x,p)
\ee
are well defined for any polynomial $F$ and thus provide meaningful
component expressions for any $F$ decomposed in formal power series in
momenta.
For $D>0$ the functional $I_D$ is analytic in $D$ in the sense
the integral of a degree-$2k$ monomial is proportional to ${D}^{d/2+k}$:
\be
I_D [F^{m(2k)} p_{m_1}...p_{m_s}] = \mu^{d+ 2k} \kappa_{d,k}
{F^{m(k)}}_{m(k)},
\ee
where $\kappa_{d,k}$ is a nonzero coefficient we do not need here,
and the integrals of odd-degree monomials are zero. Therefore, one
can define the derivatives of $I_D$ by $D$ as follows
\be \label{bint3}
I^{n}_D [F] = (\frac{\pa}{\pa D})^n \int d^d p \; \theta(-H+D) F(x,p)
\equiv \int d^d p \; \delta^{(n-1)}(-H+D) F(x,p),
\ee
that defines the distributions, corresponding to
$\delta$-function and its derivatives, $\delta^{(n-1)}(-H+D)$, for all $n>0$
(we have used the identity $\pa_{\sigma} \theta(\sigma) = \delta({\sigma})$).
For our purposes it is important that these distributions satisfy all
the standard $\delta$-function identities like
$\delta^{(n)}(\sigma) \sigma^{n+1} =0$. It becomes
clear if one calculates all the integrals in spherical coordinates with
the radial coordinate being $r=\sqrt{H}$, then the equations like
\be  \label{delf}
\delta^{(n)}(-H+\mu^2) (-H+\mu^2)^{n+1} =0
\ee
are valid because they originate from the one-dimensional identities.
Besides, the standard property
\be
0= \int d^d p \frac{\partial}{\partial p_m} \left\{ \delta^{(n)}(-H+\mu^2) F (x,p)\right\} =
\int
\delta^{(n)} (-H+\mu^2) \frac{\partial}{\partial p_m} F -
2 \delta^{(n+1)}(-H+\mu^2) p^m F \ee
holds, as it also may be checked by rewriting these integrals
in spherical coordinates. These identities allow one to
integrate by parts and are important in the subsequent check
of the wave operators selfadjointness. Given two functions $F_1$ and $F_2$
and a differential operator
$C$, we define the adjoint operator $C^\dag$ by integrating by parts, or
\be
 \int {d^d}\,p
\; F_1 \, (C \,F_2) =
\int {d^d}\,p
\; (C^{\dag} \,F_2  )\, \,F_1 \;\;\;\;\forall F_1\,,\,F_2.
\ee
In this sense, the standard properties
\be  \label{conj}
(p_n)^{\dag}=p_n \;\;,\;\;(\de^n)^{\dag}=-\de^n
\ee
are valid.
We look for an action in the form
\be \label{ans}
{\cal A}=
\int {d^d}\,x \sum\limits_{k=0}^{\infty} I_D^{k} [L^{(k)}(h_1, h_2)],
\ee
where $L^{(k)}(h_1, h_2)$ are some $(x,p)$-local expressions constructed from
$h_{1,2}$ and their $(x,p)$-derivatives.
The typical equations of motion
\be \label{eomg} E[h_{1,2}]\,(x,p)=\frac{\de {\cal A}}{\de h_{1,2}(x,p)} \ee
obtained by varying the functionals
of this sort contain some $\delta$-function pre-factors, but this peculiarity
is harmless and moreover it exhibits the fact that the action \re{ans}
is well defined in components. Indeed, the equations of motion, corresponding
to component fields $h_{1,2}^{m(s)}$ are obtained as
\be
\frac{\de {\cal A}}{\de h_{1,2}^{m(s)}(x)} =
\int {d^d}\,x'\int {d^d}\,p' \frac{\de {\cal A}}{\de h_{1,2}(x',p')}
\frac{\de  h_{1,2}(x',p')}{\de h_{1,2}^{m(s)}(x)}
=\int {d^d}\,p\; E[h_{1,2}] (x,p) \;p_{m_1}...p_{m_s},
\ee
that is well defined just because of $\delta$-function pre-factors.
Here we will use only the generating form of the equations of motions
\re{eomg}. The $\delta$-pre-factors will play an important role in checking
the invariance w.r.t. algebraic $a$ and $c$ gauge symmetries.
Let us construct the completely invariant wave equation. As the
almost wave operator $N$ \re{alm} is already $\e$-invariant, one has
to find the $a$- and $c$-invariant combination of the components
$(Nh)_1$ and $(Nh)_2$. First one finds the relation
\be \label{weyl}
\delta \left\{ (N h)_1 +\frac{H-\mu^2}{2} (Nh)_2\right\} =
\frac{(H -\mu^2)^2}{2} \left(N_{21}\, a +N_{22} \,c\right),
\ee
which holds as a consequence of the equations
\be \label{art}
\ba
{[ N_{21}, H ]} = -2 ( N_{11} + N_{22} )\\ \\
{[ N_{11}, H ]} = -2  N_{12}  \\ \\
{[ N_{22}, H ]} = -2  N_{12}.
\ea
\ee
Therefore, the equation of motion
\be \label{teq}
E[h_1,h_2]=\delta'(-H+\mu^2) \left\{ (N h)_1 +\frac{H-\mu^2}{2}
(Nh)_2\right\}, \ee
where $\delta'(-H+\mu^2)$ is a first derivative of delta-function,
is completely gauge invariant due to the identity \re{delf} with
$n=1$. It is worth rewriting this equation in terms of the unique
$a$-invariant combination
\be \label{4}
{\tilde h} =h_1-\frac{H-\mu^2}{2} h_2,
\ee
which serves also as a generating object for
Fronsdal fields since the first two traces of $\tilde h$
are linear combinations of $f_{1,2}$ as it is clear from \re{4}
and \re{5}, while the rest traces of ${\tilde h}$ are gauged away by
$c$-transformations. The rewriting
is easily done by expressing $h_1$ via $\tilde h$ and $h_2$ and
then observing that, due to \re{weyl} and transformation
laws $\delta_a {\tilde h} =0, \de_a h_2
=a$, $h_2$ does not enter the equation \re{teq} at all so that
\be \label{teq1}
\ba
\left.E[h_1,h_2]=E[{\tilde h}]=\delta'(-H+\mu^2) \left\{ (N h)_1
+\frac{H-\mu^2}{2} (Nh)_2\right\}\right|_{h_1={\tilde h}, h_2=0} =\\
\\
\delta'(-H+\mu^2) \left\{ N_{11}
+\frac{H-\mu^2}{2} N_{21}\right\}\,{\tilde h} =\\  \\
 \delta'(-H+\mu^2) \left\{AB-2BA -V_{11} + \frac{H-\mu^2}{2} (A^2
-V_{21}) \right\} {\tilde h} \equiv {\cal N} {\tilde h}. \ea  \ee
Remarkably, the wave operator
${\cal N}$ appears to be symmetric:
\be
\int {d^d}\,x \int {d^d}\,p\;
F_1 \,{\cal N} \,F_2 =
\int {d^d}\,x \int {d^d}\,p \;
\,F_2\, {\cal N} \,F_1 \;\;\;\;\forall F_1\,,\,F_2,
\ee
as may be checked by an explicit calculation involving the
integration by parts, the properties \re{conj} along with
\be
(x^m)^{\dag}=(x^m)\;,\; (\pa_m)^{\dag}=-\pa_m,
\ee
their consequences
\be
A^{\dag}=A\;\;,\;\; B^{\dag}=-B\;\;,\;\;H^{\dag} =H
\ee
etc., like
\be
(N_{11})^{\dag}=-N_{22}\;\;,\;\;
(N_{21})^{\dag}=N_{21}\;\;,\;\;(N_{12})^{\dag}=N_{12},
\ee
the commutation relations \re{art} along with ${[N_{12}, H]}=0$,
and $\de$-function identities \re{delf}.

Therefore, the invariant action is
\be   \label{sac}
\ba
{\cal A}{[h_1, h_2]} =  {\cal A}{[{\tilde h}]}=
\int {d^d}\,x \int {d^d}\,p
\;{\tilde h}\, {\cal N} \,{\tilde h} =\\ \\
\int {d^d}\,x \int {d^d}\,p\;
{\tilde h}\, {
\delta'(-H+\mu^2) \left\{AB-2BA -V_{11} + \frac{H-\mu^2}{2} (A^2
-V_{21}) \right\}}\; {\tilde h}
\ea \ee
After integration by momenta this action is equal to the
sum of Fronsdal actions for spins $s=0,1,2,3,...$ as it follows from
the analysis of gauge transformations \re{Glaws}. Indeed, we have
shown that the gauge transformations \re{Glaws} are equivalent to
the standard ones \re{frlaw} for the infinite set of integer spin
fields. The Fronsdal gauge laws \re{frlaw} for each spin-$s$ model
are known to fix the spin-$s$ action up to an overall coefficient,
provided the action does not involve higher derivatives.
Therefore, as far as the invariant of
transformations \re{Glaws}, which does not contain higher
derivatives, is constructed, it should be equal to a sum of Fronsdal
actions with some coefficients:
\be
{\cal A}[{\tilde h}]= \sum\limits_{s=0}^{\infty} \,\alpha_s\,{\cal A}_s[\vp_s],
\ee
where the map between $\tilde h$ and the collection of double traceless
$\vp_s$ is given by the formulas \re{1}, \re{5}, \re{2}, and \re{4}
(note that the second and further traces of $\tilde h$ do not enter the action
due to the $c$-symmetry, therefore only the traceless and the first trace parts of
$\tilde h$ contribute the action and these are the fields generating the infinity of
Fronsdal double-traceless tensors).
In the action \re{sac}, the $V$-terms are the $\La$-proportional mass-like
terms mentioned in the introduction (note that $V$ does not include
$x$-derivatives).

Summing up, we have presented the new generating  formulation for
free massless higher spin dynamics. It is constructed in terms of
two fields $h_{1,2}(x,p)$ subject to simple gauge transformations
\re{Glaws} while the invariant action \re{sac} has the form of an
integral localized on the "constraint surface" $H-\mu^2 =0$ in the
cotangent bundle of $AdS_d$ space. The reduction to the constraint surface
serves for removing the traces of tensor coefficients of the fields
$h_{1,2}(x,p)$, leaving just the infinite collection of double traceless
Fronsdal fields.

The appearance of the cotangent bundle and the constraint surface is
natural in the spirit of the author's paper \cite{Segal:2000ke}, where
an analysis of a $d$-dimensional point particle dynamics in general
background fields had been undertaken. The backgrounds are parametrized
with an arbitrary function ${\cal H}(x,p)$ which sets up the dynamics
on the constraint surface ${\cal H}(x,p) \approx 0$.  It was shown that the
gauge transformations for $h_1$ may be interpreted as the
linearization, around the "vacuum" ${\cal H}(x,p)= -\mu^2 +p^2$, of
nonlinear "generalized equivalence transformations" for $H$.
The localization of the dynamics on the surface ${\cal H}=0$ exhibited by
$\delta$-function pre-factors in \re{sac} is natural in the spirit of
this interpretation of $\cal H$. On the other hand,
it was observed that, apart from $h_1$,
additional degrees of freedom are needed to
get Fronsdal fields. The second function $h_2$ appears to be
a good complement as it is demonstrated in the present paper.

We think the new formulation may provide a good starting point for
analyzing the interaction problem. It is interesting to find a geometry
characterized by two functions, say $H(x,p)$ and $G(x,p)$,
and some covariance group and possessing a natural invariant
functional ${\cal A}[H,G]$
in such a way that the expansion of the covariance transformations
around some vacuum state gives the gauge laws \re{Glaws} for fluctuations
$h_{1,2}$, while the linearization of the invariant ${\cal A}[H,G]$
gives the action \re{sac}. Then ${\cal A}[H,G]$ may be considered as a
consistent higher spin action. In the spirit of this program, our
formulation seems preferable as compared to another
generating models \cite{klishevich}-\cite{Bellon:1987ki}
which use auxiliary operators of creation and annihilation
that seems harder to deform to the nonlinear case.

It is worth noting that an analogous formulation  for free
$d=4$ supersymmetric theories had been constructed in \cite{gks},
where the "unified" higher spin action is built out of a few scalar
fields living on the manifold ${\cal M}^{7|4}$ being the extension of
the $AdS_4$ superspace by a 3-dimensional space-like hyperboloid.
A reformulation in terms of the extension by a
time-like hyperboloid, ${\bar M}^{7|4}$ is also possible
\cite{sibir}. The superspace ${\bar M}^{7|4}$ may be viewed as a
constraint surface in the extended phase space of
$N=1,d=4$ $AdS$ superparticle, that strengthen
the analogy to the purely boson construction of the present paper.

It is likely this resemblance is not accidental  and manifests the
existence  of the underlying nonlinear "higher spin geometry" either for
SUSY or non-SUSY case.

\section*{Acknowledgement}
The work is supported by grants RFBR 99-02-16207 and RFBR 00-15-96566.

\end{document}